\documentclass{article}

\usepackage{authblk}
\usepackage{cite}
\usepackage{graphicx}
\usepackage{amsmath,amssymb}
\usepackage{array}
\usepackage{subfig}

\def\wideimwidth{17cm}
\def\midimwidth{11cm}
\def\narrowimwidth{9cm}

\begin{document}

\title{Simultaneous Reconstruction of Emission and Attenuation in Passive Gamma Emission Tomography of Spent Nuclear Fuel}

\author[1,2,$\dagger$]{R.~Backholm}
\author[2]{T.~A.~Bubba}
\author[1,$\ddagger$]{C.~B\'{e}langer-Champagne}
\author[3]{T.~Helin}
\author[1]{P.~Dendooven}
\author[2]{S.~Siltanen}
\affil[1]{Helsinki Institute of Physics, University of Helsinki, Finland}
\affil[2]{Department of Mathematics and Statistics, University of Helsinki, Finland}
\affil[3]{School of Engineering Science, LUT University, Finland}

\footnotetext[2]{Currently with STUK, Finland}
\footnotetext[3]{Currently with TRIUMF, Canada}

\maketitle

\begin{abstract}
The International Atomic Energy Agency (IAEA) has recently approved passive gamma emission tomography (PGET) as a method for inspecting spent nuclear fuel assemblies (SFAs), an important aspect of international nuclear safeguards which aim at preventing the proliferation of nuclear weapons. The PGET instrument is essentially a single photon emission computed tomography (SPECT) system that allows the reconstruction of axial cross-sections of the emission map of the SFA. The fuel material heavily self-attenuates its gamma-ray emissions, so that correctly accounting for the attenuation is a critical factor in producing accurate images. Due to the nature of the inspections, it is desirable to use as little \textit{a priori} information as possible about the fuel, including the attenuation map, in the reconstruction process. Current reconstruction methods either do not correct for attenuation, assume a uniform attenuation throughout the fuel assembly, or assume an attenuation map based on an initial filtered back projection (FBP) reconstruction. Here, we propose a method to simultaneously reconstruct the emission and attenuation maps by formulating the reconstruction as a constrained minimization problem with a least squares data fidelity term and regularization terms. Using simulated data, we compare the proposed method to FBP, showing that our approach produces significantly better reconstructions by various numerical metrics and a much better classification of spent, missing, and fresh fuel rods.
\end{abstract}

\newpage

\section{Introduction}

As part of an effort to deter the proliferation of nuclear weapons, various technical measures referred to as "safeguards" are used to verify the declarations made by the signatories to the Treaty on the Non-Proliferation of Nuclear Weapons about their nuclear material and activities \cite{UN1970}. The monitoring of spent fuel assemblies (SFAs) from nuclear power plants (NPPs) is an important task within these safeguards, aiming at detecting any eventual diversion of spent nuclear fuel for non-declared purposes. Ideally, a single fuel pin missing from an SFA should be detected. For any safeguards investigation of SFAs, it is important to use a minimum amount of \textit{a priori} information on the SFA under study, in order to avoid biasing and potentially misleading the investigation.

Each non-nuclear-weapon State Party to the Treaty is required to conclude a safeguards agreement with the International Atomic Energy Agency (IAEA). As such, safeguards activities are largely monitored and coordinated by the IAEA.

Since the 1980s, IAEA has developed, in collaboration with some of its Member States, gamma ray emission tomography (GET) for imaging SFAs \cite{Levai1993}. GET was deemed attractive for detecting partial defects (part of the fuel of an SFA missing) and verifying the integrity of the SFA because it has the potential to directly image the spatial distribution of the active material and the relative locations of the pins in the SFA in a non-destructive way. This effort has culminated at the end of 2017 in IAEA approval to use the PGET instrument (Passive Gamma Emission Tomography) \cite{White2017, White2018, Mayorov2017} in inspections.

The PGET instrument is able to reliably identify single missing or replaced pins in WWER-440, BWR and PWR SFAs with burnups in the range of 5.7–-57.8 GWd/tU and cooling times from 1.9-–26.6 years \cite{White2017, White2018, Mayorov2017}. Images reconstructed using gamma ray energies higher than those of \textsuperscript{137}Cs (\textgreater700~keV) have better water-to-fuel contrast in fuel cooled for up to 20 years and thus have the most potential for missing fuel pin detection \cite{BelangerChampagne2019}. Following these first results, IAEA has expressed the need for new image reconstruction and processing methods for a more accurate assessment of the locations and count of missing pins and a more accurate calculation of the relative radioactivity levels of individual pins.

An SFA is a challenging object for tomographic imaging as it contains materials with very different emission and attenuation properties: strongly attenuating and emitting spent nuclear fuel (commonly uranium dioxide) and less attenuating material with zero emission (water or air). When considering diversion scenarios beyond missing pins, one could also consider, e.g., pins replaced with fresh nuclear fuel (strongly attenuating and zero emission) or with activated materials other than nuclear fuel (moderately attenuating and high emission). 

An SFA consists of a regular lattice of about 100 to 300 fuel pins depending on fuel type, with most often one or several empty lattice positions (so-called water channels). Attenuation of the 662~keV gamma rays from \textsuperscript{137}Cs between the center and the edge of an SFA is of the order of a factor 100. This high gamma ray attenuation combined with the heterogeneous nature of SFAs makes detailed attenuation information essential for producing very realistic images.

Acquiring such attenuation information by means of a separate imaging procedure such as high energy CT is not practical from an operational point of view. Information on the geometry of the SFA, e.g., provided by the NPP, can in principle be used to obtain detailed information on attenuation. However, the requirement to make use of as little \textit{a priori} information as possible precludes this.

On the other hand, partial defect detection does not necessarily require images with accurate (relative) intensities: it is more important to have good contrast between emitting and non-emitting regions. In practice, however, images with more accurate (relative) intensities typically have better contrast.

The development of GET for spent fuel verification has largely been conducted under the IAEA Support Program projects JNT 1510 and JNT 1955 (phase I). The JNT 1955 project \cite{Smith2016, JacobssonSvard2017} and related work \cite{JacobssonSvard2015} used simulated data and investigated both analytic (filtered back-projection (FBP) wihout and with \textit{a posteriori} attenuation correction) and algebraic image reconstruction techniques (which combined the Additive Simultaneous Iterative Reconstruction Technique with homogeneous attenuation information throughout the area covered by the SFA).

The PGET instrument recently approved by IAEA for inspections resulted from the JNT 1510 project \cite{Honkamaa2014}. In its implementation as approved for SFA inspections, the image from an initial FBP without attenuation correction (so without \textit{a priori} knowledge) is used to determine (assumed) pin locations and fuel assembly location. This information is then used to construct a heterogeneous attenuation map that is included in a second image reconstruction using the Novikov inversion formula \cite{Kunyansky2001} and resulting in the final image \cite{White2017}.

In an effort to get closer to the goal of not using any \textit{a priori} information on the materials and geometry in GET for SFA safeguards, we investigated an approach for the simultaneous reconstruction of emission and attenuation. In the context of medical single photon emission computed tomography (SPECT), the simultaneous reconstruction from emission data has been a topic of research since the late 1970s \cite{Censor1979}. For a recent extensive review, we refer to \cite{Berker2016}. Here, we mention only some iterative methods used for simultaneous reconstruction in the case of arbitrary attenuation maps \cite{Censor1979, Manglos1993, Dicken1999, Nuyts1999, Hawkins1999, Krol2001, Gourion2002, Krol2002}.

The approach we propose is close  to the ones in \cite{Dicken1999} and \cite{Gourion2002}. Indeed, we also formulate the reconstruction as a minimization problem with a least squares (LSQ) data fidelity term and regularization terms. However, our choice for the regularizers and the minimization algorithm differ, and we also use linear bounds that are specific for the application.

In particular, we investigate two regularization terms. The smoothness prior has been extensively used in linear tomography problems yielding satisfactory results given that it is computationally efficient and simple to implement. On the other hand, the geometry aware prior, to the best of our knowledge, has not been proposed before and is tailored for this specific application. Compared to the smoothness prior, the geometry aware prior maintains the computational efficiency, but improves the reconstruction at the small cost of reasonable assumptions about the fuel assembly geometry, available, for instance, from an initial FBP reconstruction.

This paper is organized as follows. In Section~\ref{sec:methods} we describe the PGET instrument, introduce the discrete measurement model along with the simulation of data and the minimization problem. Results are presented in Section~\ref{sec:results} and discussed in Section~\ref{sec:discussion}. We draw some conclusions and indicate future perspectives in Section~\ref{sec:conclusions}.

\section{Methods}
\label{sec:methods}

\subsection{Measurement with the PGET instrument}

\begin{figure}[!t]

\centerline{\includegraphics[width=\wideimwidth]{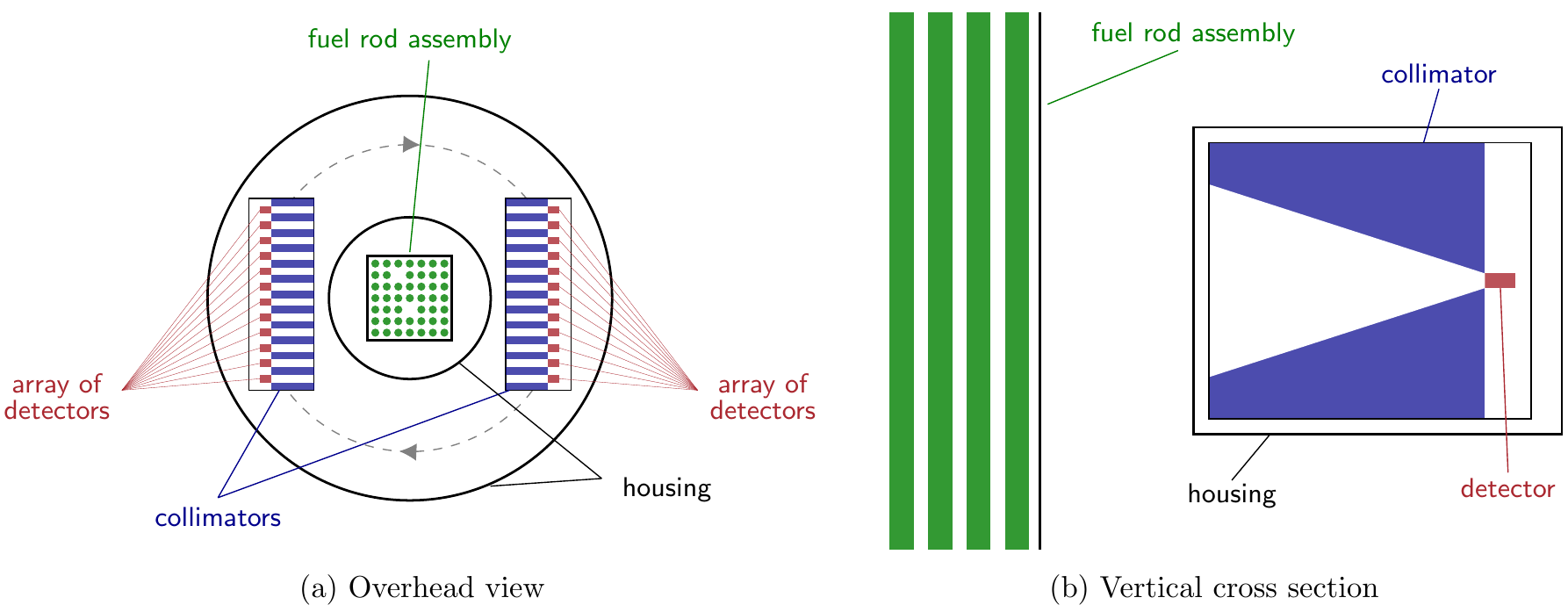}}

\captionsetup{width=\wideimwidth}
\caption{Simplified schematics of the PGET instrument. (a) Two detector banks on opposite sides of an SFA being measured. (b) Collimator slit profile and the location of the detectors with respect to the fuel rods.}
\label{fig:schem}

\end{figure}

The PGET safeguards instrument measures the gamma ray emissions from the nuclear fuel.  It is a 1D SPECT system, using a 1D linear collimator in front of a 1D array of gamma ray detectors. This geometry allows for the reconstruction of 2D cross-section images of the fuel. A simplified schematic representation of the instrument is shown in Figure~\ref{fig:schem}.  The PGET is made up of 2 detector banks with 87 CZT gamma ray detectors behind a tungsten collimator in each bank mounted on a plate inside a water-tight enclosure. During the course of each measurement, that plate rotates 360 degrees to measure data projections around the whole fuel assembly.

The CZT detectors have dimensions $2\text{\ mm}\times4.8\text{\ mm}\times4.8\text{\ mm}$. The detector and collimator pitch in each bank is 4~mm, and the position of the 2 banks on opposite sides of the rotating plate are offset by 2~mm so that the detectors of the banks can be interleaved to achieve an effective detector spacing of 2~mm.

The tungsten linear collimator slits are 10~cm deep, 1.5~mm wide, and taper from 70~mm tall at the front to 5~mm at the back. On all other sides except for the collimator opening, the detectors are shielded by at least 2~cm of tungsten. The water-tight enclosure is made out of 3-mm-thick stainless steel plating.

For each measured data projection, each CZT detector records the number of counts above 4 user-determined gamma-ray energy thresholds over a user-determined measurement time.  These are used to calculate the number of gamma-ray counts in broad energy windows defined between consecutive energy thresholds.  The thresholds are typically selected to enhance the contribution of a single gamma-ray emitting isotope (\textsuperscript{137}Cs, \textsuperscript{154}Eu) in each window. More detailed descriptions of the PGET instrument can be found in \cite{White2018, BelangerChampagne2019, Honkamaa2014}.

The PGET instrument geometry was pixelized in order to create the measurement model used in our proposed image reconstruction algorithm.

\subsection{Discrete measurement model}
\label{subsection:discmodel}

\begin{figure}[!t]
\centering

\includegraphics[width=\midimwidth]{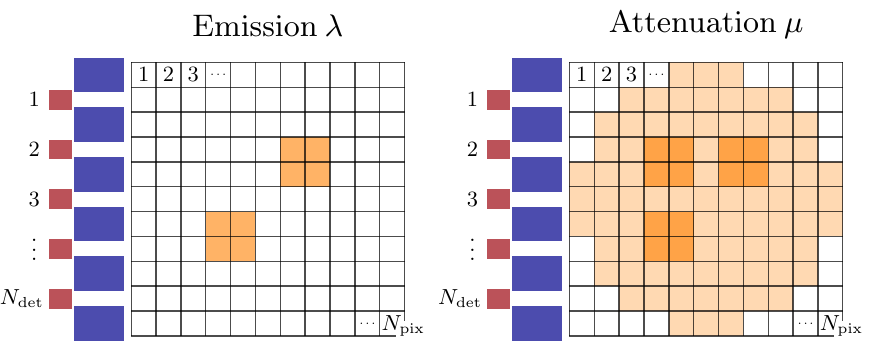}

\captionsetup{width=\midimwidth}
\caption{Scaled-down example of a discrete emission map $\lambda$ (left), a discrete attenuation map $\mu$ (right), and the pixel indexing. Next to the maps is the detector array at the position corresponding to measurement angle zero. The collimators are in blue, and the detectors, shown with their indexing, are in red.}
\label{fig:mapimages}

\end{figure}

We assume that the volume contributing to the measurement is uniform in its emission and attenuation along the direction of the fuel rods. Further, we represent the volume by its 2D axial cross-section, which we divide into an $n$ by $n$ grid of pixels indexed from $1$ to \mbox{$N_{\text{pix}}=n^{2}$}.  We denote by \mbox{$\lambda=(\lambda_{1},\dotsc,\lambda_{N_{\text{pix}}})\in\mathbb{R}^{N_{\text{pix}}}_+$} the vector of the emission values of the cross-section, and likewise by $\mu=(\mu_{1},\dotsc,\mu_{N_{\text{pix}}})\in\mathbb{R}^{N_{\text{pix}}}_+$ the attenuation values. Here  $\mathbb{R}_+$ is the set of non-negative real numbers. A scaled-down example of these discrete cross-section maps and the pixel indexing can be seen in Figure~\ref{fig:mapimages}.

Only those elements of $\lambda$ and $\mu$ that correspond to pixels inside the maximal circular disk contained in the $n\times n$ images are within the region of interest in the measurement and thus are actually variables. This disk can be seen in the attenuation map in Figure~\ref{fig:mapimages}. The elements outside the disk are always set to zero. This is easily implemented in practice and we ignore this consideration in the rest of the article to simplify the following descriptions.

As in \cite{Seppanen200}, we first describe how the measurements are formed in the case that the detector array, with $N_{\text{det}}$ detectors, is located to the side of the cross-section images, as seen in Figure~\ref{fig:mapimages}. This detector position is considered here to be the zero measurement angle. The measurements at other angles are then easily computed using this zero angle setup and rotating the contents of the emission and attenuation images.

The forward projection at the zero measurement angle can be expressed in the form \mbox{$F_{0}(\lambda,\mu)=H_{0}(\mu)\lambda$}, where $H_{0}(\mu)$ is a $N_{\text{det}}\times N_{\text{pix}}$ matrix depending on $\mu$. The element $H_{0}(\mu)_{i,p}$, which is the coefficient by which the emission value of pixel $p$, that is $\lambda_{p}$, contributes to the measurement at detector $i$, can be expressed as
\begin{align}
H_{0}(\mu)_{i,p}=r_{i,p}\exp\left(-c_{i,p}d_{i,p}^{T}\mu\right).\label{eq:zeroangle}
\end{align}

Here $r_{i,p}\in\mathbb{R}_+$ is the spatial response of pixel $p$ with regard to detector $i$, namely, it expresses the probability that photons emitted isotropically in the volume represented by pixel $p$ propagate towards the visible part of the detector $i$. How this is determined is described in detail below.

\begin{figure}[!t]
\centering

\includegraphics[width=\midimwidth]{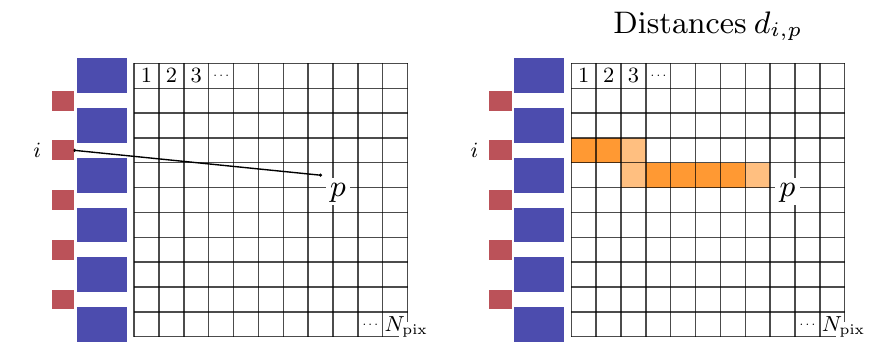}

\captionsetup{width=\midimwidth}
\caption{Line from the center of pixel $p$ to the center of detector $i$ (left), and $d_{i,p}$ which tells for every pixel in the grid the length that the aforementioned line travels inside that pixel (right). }
\label{fig:distimages}

\end{figure}

The $q$th component of the vector $d_{i,p}\in\mathbb{R}^{N_{\text{pix}}}_+$ is the distance that the line connecting the center of pixel $p$ and the center of detector $i$ travels inside pixel $q$. This is illustrated in Figure~\ref{fig:distimages}. The product $d_{i,p}^{T}\mu$ can be understood as a line integral of $\mu$ along the line from pixel $p$ to detector $i$.

The term $c_{i,p}>1$ is a correction factor for the distances $d_{i,p}$. This is to take into account the fact that the distance traveled by photons emitted from the volume represented by pixel $p$ (i.e. the vertical extent of the SFA seen through the collimator slits) is usually longer than the distance from the center of pixel $p$ to detector $i$. How this is formed is described in detail below.

\begin{figure}[!t]
\centering

\includegraphics[width=\midimwidth]{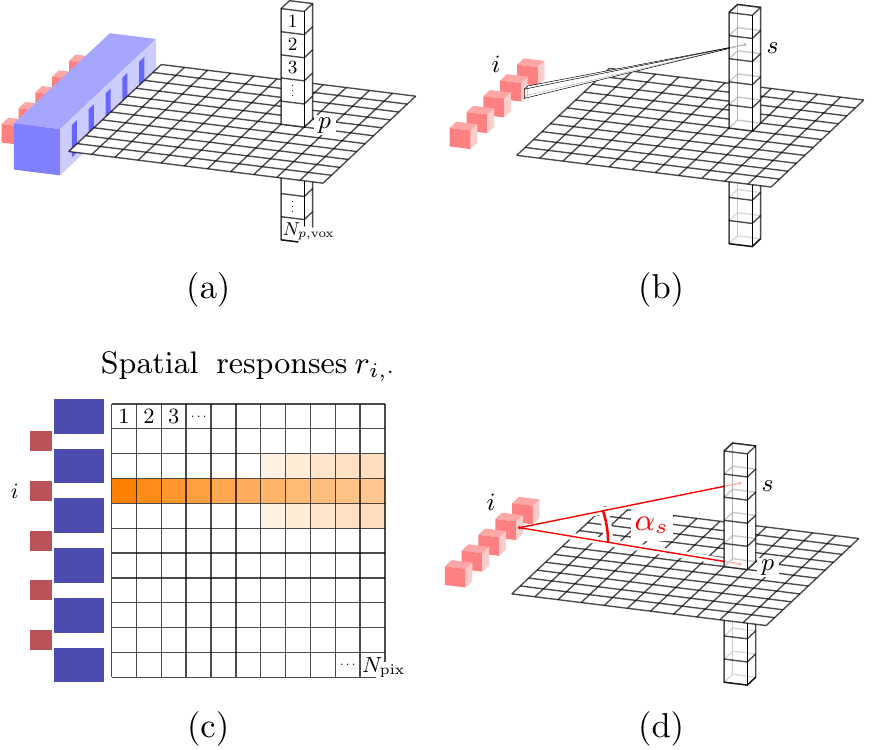}

\captionsetup{width=\midimwidth}
\caption{(a) The volume that pixel $p$ represents is divided into voxels. (b) The cone spanned by the visible part of detector $i$ from voxel $s$ defines a solid angle. (c) Spatial responses $r_{i,p}$ of detector $i$ for all the pixels $p$ in the grid. (d) The angle $\alpha_s$ between the line from pixel $p$ to detector $i$ and the line from voxel $s$ to detector $i$.}
\label{fig:voximages}

\end{figure}

The spatial response $r_{i,p}$ is computed similarly to \cite{Thierry1999}. First the volume that pixel $p$ represents is divided into voxels, indexed from $1$ to $N_{p,\text{vox}}$, as seen in Figure~\ref{fig:voximages}(a). Now the spatial response of each voxel $s$, denoted by $r_{s}^{3D}$ (we drop the dependence on $i$ and $p$ from the notation for simplicity), is just the probability that a photon emitted from the center of voxel $s$ starts off towards the visible part of detector $i$. This is equal to the solid angle spanned at the center of voxel $s$ by the visible part of detector $i$ divided by $4\pi$ (Figure~\ref{fig:voximages}(b)). The spatial response $r_{i,p}$ is then just the average of the spatial responses of individual voxels (Figure~\ref{fig:voximages}(c)):
\begin{align}
r_{i,p}=\frac{1}{N_{p,\text{vox}}}\sum_{s=1}^{N_{p,\text{vox}}}r_{s}^{3D}.
\end{align}

For the correction factor $c_{i,p}$ consider the angle $\alpha_{s}$ between the line from pixel $p$ to detector $i$ and the line from voxel $s$ to detector $i$ as illustrated in Figure~\ref{fig:voximages}(d). Multiplying the length of the former line by $1/\text{cos}\alpha_{s}$ gives the length of the latter line. The correction factor $c_{i,p}$ is now the weighted average of the factors $1/\text{cos}\alpha_{s}$ with the spatial responses $r_{s}^{3D}$ used as weights: 
\begin{align}
c_{i,p}=\frac{N_{p,\text{vox}}}{r_{i,p}}\sum_{s=1}^{N_{p,\text{vox}}}\frac{r_{s}^{3D}}{\cos\alpha_{s}}.
\end{align}

If the image resolution $n\times n$ is low, then it is advantageous to compute the spatial responses $r$ and the correction factors $c$ using a higher resolution and then downsample them to $n\times n$ by averaging.

The forward projection at an arbitrary angle $\phi$ can now be expressed as \mbox{$F_{\phi}(\lambda,\mu)=H_{\phi}(\mu)R_{\phi}\lambda$}, where $R_{\phi}$ is a $N_{\text{pix}}\times N_{\text{pix}}$ matrix that rotates the contents of the cross-section images by angle $\phi$ using bilinear interpolation, and $H_{\phi}(\mu)$ is a $N_{\text{pix}}\times N_{\text{pix}}$ matrix defined similarly to $H_{0}(\mu)$ in (\ref{eq:zeroangle}):
\begin{align}
H_{\phi}(\mu)_{i,p}=r_{i,p}\exp\left(-c_{i,p}d_{i,p}^{T}R_{\phi}\mu\right).
\end{align}

Finally, the whole forward projection with measurement angles $\phi_{1},\dotsc,\phi_{N_{\text{ang}}}$ can be expressed as $F(\lambda,\mu)=H(\mu)\lambda$, where $H(\mu)$ is the $N_{\text{det}}\cdot N_{\text{ang}}\times N_{\text{pix}}$ system matrix 
\begin{align}
H(\mu)=\left[\begin{matrix}H_{\phi_{1}}(\mu)R_{\phi_{1}}\\
\vdots\\
H_{\phi_{N_{\text{ang}}}}(\mu)R_{\phi_{N_{\text{ang}}}}\label{eq:Hcomp}
\end{matrix}\right].
\end{align}

\subsection{Simulation of data}\label{subsection:simulation}

Recovering both the attenuation and emission simultaneously is a nonlinear and ill-posed inverse problem. Therefore, it is important to avoid the so-called {\it inverse crime} \cite[Ch. 2.3]{Mueller2012}. In other words, the simulated data should not be produced by exactly the  same computational model that is used in the reconstruction algorithm, namely the model in Section \ref{subsection:discmodel}.

For this purpose, we briefly introduce another model, which is essentially a fully 3D version of the previous one. Instead of considering only a cross-section and dividing that into pixels, we divide the whole volume that can be seen by the detectors through the collimator slits into voxels, indexed $1,\dotsc,N_{\text{vox}}$. 

This 3D model can be described in much the same terms as the previous 2D one. We use notation with tilde for the concepts related to the 3D model. Let $\tilde{\lambda},\tilde{\mu}\in\mathbb{R}_{+}^{N_{\text{vox}}}$ denote the discrete emission and attenuation maps of the volume. Now the 3D forward projection at angle $\phi$ can be expressed as \mbox{$\tilde{F}_\phi(\tilde{\lambda},\tilde{\mu})=\tilde{H}_{\phi}(\tilde{\mu})\tilde{R}_\phi\tilde{\lambda}$}, where $\tilde{H}_{\phi}(\tilde{\mu})$ is a $N_{\text{det}}\times N_{\text{vox}}$ matrix depending on $\tilde{\mu}$ and $\tilde{R}_{\phi}$ is a $N_{\text{vox}}\times N_{\text{vox}}$ matrix that rotates the volume by $\phi$ degrees. The element $\tilde{H}_{\phi}(\tilde{\mu})_{i,s}$, which is the coefficient by which the emission value of voxel $s$, that is $\tilde{\lambda}_{s}$, contributes to the measurement at detector $i$, can be expressed as 
\begin{align}
\tilde{H}_{\phi}(\tilde{\mu})_{i,s}=\tilde{r}_{i,s}\exp\left(-\tilde{d}_{i,s}^{\:T}\tilde{R}_{\phi}\tilde{\mu}\right).
\end{align}
The term $\tilde{r}_{i,s}$ is the 3D spatial response. This is actually the same spatial response $r_s^{3D}$ that was used to compute the 2D spatial response in Section~\ref{subsection:discmodel}. The $u$th element of $\tilde{d}_{i,s}$ is the distance that the line connecting the center of voxel $s$ and the center of detector $i$ travels inside voxel $u$.

As in the 2D case, all the matrices $\tilde{H}_{\phi}(\tilde{\mu})$ can be composed so that the whole forward projection can be expressed as \mbox{$\tilde{F}(\tilde{\lambda},\tilde{\mu})=\tilde{H}(\tilde{\mu})\tilde{\lambda}$}, where the system matrix $\tilde{H}(\tilde{\mu})$ is similar to (\ref{eq:Hcomp}).

\subsection{Minimization problem}

We formulate the reconstruction from a measurement $m\in\mathbb{R}^{N_{\text{det}}\cdot N_{\text{ang}}}$ as a constrained minimization problem with a LSQ data fidelity term and regularization terms $P_i$:
\begin{align}
\label{eq:minprob}
\begin{split}
\text{min}_{(\lambda,\mu)\in\mathbb{R}^{2N_{\text{pix}}}}&\left\{\left\Vert F\left(\lambda,\mu\right)-m\right\Vert _{2}^{2}+\sum_i\alpha_i P_i\left(\lambda,\mu\right)\right\} \\ 
&\text{subject\:to}\quad A\left[
\begin{matrix}\lambda \\
\mu
\end{matrix}\right]
\leq b.
\end{split}
\end{align}

The purpose of the regularization terms is to compensate for the incomplete data by incorporating {\it a priori} knowledge about the unknowns in the reconstruction. Regularized inversion is robust against modelling errors and measurement noise. For more information on regularization of nonlinear ill-posed inverse problems, see \cite{Mueller2012, Schuster2012}. 

The regularization parameters $\alpha_i$ balance the effect of the data fidelity term and the regularization terms. Matrix $A$ and vector $b$ are such that the inequality in (\ref{eq:minprob}), understood to hold componentwise, defines a convex set.

The data fidelity term $\left\Vert F\left(\lambda,\mu\right)-m\right\Vert _{2}^{2}$, if seen only as function of emission $\lambda$, is convex, but as function of attenuation $\mu$, or both $\lambda$ and $\mu$, it is non-convex. It is a smooth function in all cases.

\subsubsection{Regularization terms}

We use two different choices for the regularization terms. In both cases the terms are convex quadratics.

The first choice, called here \textbf{the smoothness prior}, predisposes the algorithm toward reconstructions that are smooth in the sense that the changes in the emission and attenuation images are gradual when moving from one pixel to the next. It has the form
\begin{align}
\alpha_{\lambda}\left\Vert L\lambda\right\Vert _{2}^{2}+\alpha_{\mu}\left\Vert L\mu\right\Vert _{2}^{2},\label{eq:smoothess}
\end{align}
where $L$ is the discrete Laplace operator, that is, a two dimensional convolution with the kernel
\begin{align}
\text{ker}_L=\left[\begin{array}{rrr}
0 &  1 & 0\\
1 & -4 &\:\:\: 1\\
0 &  1 & 0
\end{array}\right].
\end{align}
The convolution is computed only at the points where the non-zero elements of the kernel stay inside the disk of variables.

The other choice for the penalty terms, called \textbf{the geometry aware prior}, assumes that the positions and the diameters of the possible rods, whether they are actually present or not, are known. In practice this information can be gained, for example, by identifying the assembly type and its position from an FBP reconstruction.

\begin{figure}[!t]
\centering

\includegraphics[width=\narrowimwidth]{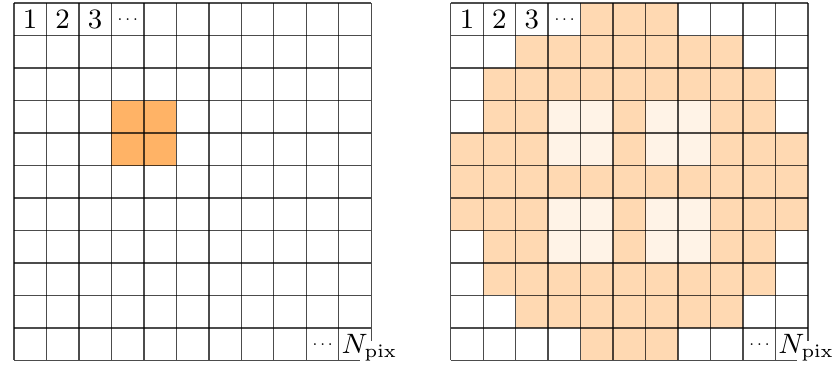}

\captionsetup{width=\narrowimwidth}
\caption{Examples of the basis images used by the geometry aware prior in the scaled-down setting with four rod positions. On the left there
is $r_{1}$, an image containing only one of the rods, and on the right there is the water image $w$. Due to the low resolution and the round shape of the rods, all the rod pixels are partly water, which is why the water image has non-zero values in the rod pixels. This is also the case for the edge pixels of the fuel rods in the full-scale setting.}
\label{fig:basis-images}

\end{figure}

Let $r_{i}\in\mathbb{R}^{N_{\text{pix}}}$, $1\leq i\leq N_{\text{rod}}$, be vectorized images each displaying one of the $N_{\text{rod}}$ rods and nothing else. In addition let $w\in\mathbb{R}^{N_{\text{pix}}}$ be a vectorized image of water outside the rod positions. A scaled-down example of these images can be seen in Figure~\ref{fig:basis-images}.

We wish to assert that our emission reconstruction is close to being a linear combination of the rod vectors $r_{i}$, i.e., that it is close to the subspace \mbox{$S_{\lambda}=\text{span}\left(r_{1},\dotsc,r_{N_{\text{rod}}}\right)$}. Similarly we wish that the attenuation reconstruction is close to being a linear combination of the rods $r_{i}$ and the water vector $w$, meaning it is close to \mbox{$S_{\mu}=\text{span}\left(r_{1},\dotsc,r_{N_{\text{rod}}},w\right)$}.

To achieve this, we define the matrices
\begin{align}
B_{\lambda}=\left[\begin{matrix}r_{1} & \dotsb & r_{N_{\text{rod}}}\end{matrix}\right]\:\text{and}\: B_{\mu}=\left[\begin{matrix}r_{1} & \dotsb & r_{N_{\text{rod}}} & w\end{matrix}\right].
\end{align}
The expression \mbox{$P_{\lambda}=I-B_{\lambda}(B_{\lambda}^{T}B_{\lambda})^{-1}B_{\lambda}^{T}$}, where $I$ is the identity matrix, is the projection onto the orthogonal complement of $S_{\lambda}$. Define $P_{\mu}$ similarly. The geometry aware prior then has the form
\begin{align}
\alpha_{\lambda}\left\Vert P_{\lambda}\lambda\right\Vert _{2}^{2}+\alpha_{\mu}\left\Vert P_{\mu}\mu\right\Vert _{2}^{2}.\label{eq:subspace}
\end{align}

\subsubsection{Bounds}

\begin{figure}[!t]
\centering

\includegraphics[width=\narrowimwidth]{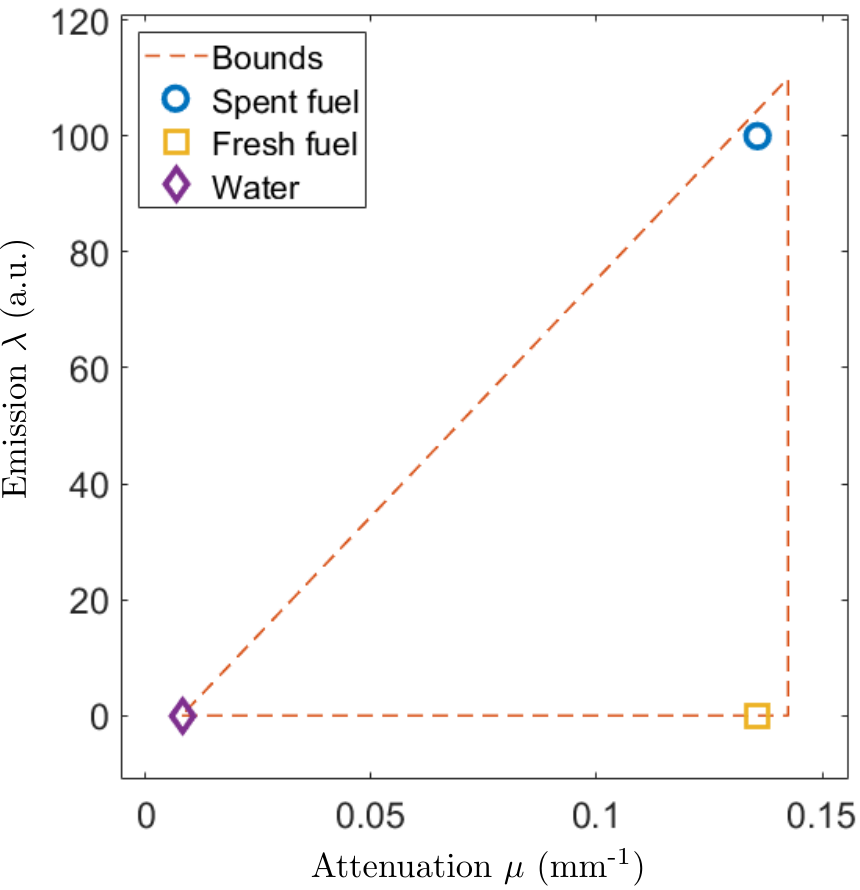}

\captionsetup{width=\narrowimwidth}
\caption{The linear bounds illustrated in the emission-attenuation-plane along with points that correspond to spent fuel (high emission, high attenuation), fresh fuel (no emission, high attenuation) and water (no emission, low attenuation). The values inside the triangle are allowed by the bounds. The triangle is slightly larger than necessary to allow the three materials mentioned, which is to simulate error from estimating the bounds. The attenuation values shown are the linear attenuation coefficients (mm\textsuperscript{-1}) of water and UO\textsubscript{2} for 662 keV gamma-rays from \textsuperscript{137}Cs. The emission values are arbitrary.}
\label{fig:bounds}

\end{figure}

The linear bounds that are used can be described as applying equally to all pairs of pixel values $(\lambda_{p},\mu_{p})$, that is, to all pairs of emission and attenuation values from the same pixel. Hence the bounds can be visualized in the emission-attenuation-plane, where they form a triangle, as seen in Figure~\ref{fig:bounds}. The values inside the triangle are feasible.

The bounds allow, in particular, the three materials relevant to us in this study: water (no emission, low attenuation), spent fuel rod (high emission, high attenuation) and fresh fuel rod (no emission, high attenuation), but they exclude the physically unlikely case of a material with high emission and low attenuation.

The triangle bounds can be described by giving an upper bound for emission values and both upper and lower bounds for attenuation values as these determine the three vertices of the triangle (the lower bound for emission is assumed to be zero). Some ways of estimating these upper and lower bounds are required in practice and this is discussed briefly in Section~\ref{sec:discussion}. Here we simply modify the true upper and lower bounds, as described in Section~\ref{sec:results}, to simulate error in estimating these values.

\subsubsection{Minimization algorithm}

The regularization terms that we use are such that the functional being minimized in (\ref{eq:minprob}) can be naturally written as non-linear LSQ term
\begin{align}
\left\Vert r(\lambda,\mu)\right\Vert _{2}^{2}:=\left\Vert \begin{matrix}F(\lambda,\mu)-m\\
 \sqrt{\alpha_\lambda}M_\lambda\lambda\\
  \sqrt{\alpha_\mu}M_\mu\mu
\end{matrix}\right\Vert _{2}^{2},\label{eq:resid}
\end{align}
where the matrices $M_\lambda$ and $M_\mu$ depend on the choice of the penalty: they are either the discrete Laplace operator $L$ from (\ref{eq:smoothess}) or the projection matrices $P_\lambda$ and $P_\mu$ from (\ref{eq:subspace}). We exploit this formulation of the problem and use a minimization method that is similar to the Levenberg-Marquardt algorithm (LMA) as  described in \cite{Kelley1999}.

Denote here by $x$ the combination of the emission and attenuation vectors, that is, $x=\left[\begin{matrix}\lambda^{T} & \mu^{T}\end{matrix}\right]^{T}$, and write $r(x)$ for the residual $r(\lambda,\mu)$ in (\ref{eq:resid}). The LMA is a method of \textit{unconstrained} optimization. At each iteration $k$, it minimizes, with regard to the next step $x_{\text{step}}$, a linear LSQ term that results from linearizing the residual $r(x)$ at the current iterate $x^{(k)}$ and from adding a regularization term:
\begin{align}
\left\Vert \begin{matrix}\left[\begin{array}{c}
J_{r}(x^{(k)})\\
\sqrt{\beta^{(k)}}I
\end{array}\right]x_{\text{step}}+\left[\begin{array}{c}
r(x^{(k)})\\
0
\end{array}\right]\end{matrix}\right\Vert _{2}^{2},\label{eq:lsq}
\end{align}
Here $J_{r}(x^{(k)})$ is the Jacobian matrix of the residual $r(x)$, $I$ is the $2N_{\text{pix}}\times 2N_{\text{pix}}$ identity matrix, and $\beta^{(k)}$ is the LM parameter modified at each step.

Differing from the usual LMA, we minimize (\ref{eq:lsq}) using linear constraints that keep the next iterate \mbox{$x^{(k+1)}=x^{(k)}+x_{\text{step}}$} feasible: 
\begin{align}
Ax_{\text{step}}\leq b-Ax^{(k)}.
\end{align}
This minimization is done using the scaled gradient projection (SGP) method \cite{Bonettini2009}, where we use for scaling the inverse of the diagonal matrix that has the same diagonal as
\begin{align}
2J_{r}(x^{(k)})^{T}J_{r}(x^{(k)})+2\beta^{(k)}I,
\end{align}
which is the Hessian of (\ref{eq:lsq}) with regards to $x_{\text{step}}$.

The Jacobian matrix $J_{r}(x^{(k)})$ is computed analytically, namely, based on an exact expression for it.

\section{Results}
\label{sec:results}

The sinogram data used in this study was simulated with the 3D model described in Section \ref{subsection:simulation}. The fuel assembly phantoms used consist of $660\times660\times 639$ voxels. Each axial column of voxels had uniform emission and attenuation properties. The measurements were simulated at 672 detector points per angle and then interpolated down to 167 detectors. Finally, Gaussian white noise, with a standard deviation of 2\% of the maximum value of all the measurements, was added to form the data used in the reconstruction process.

\begin{figure}[!t]

\centerline{\includegraphics[width=\wideimwidth]{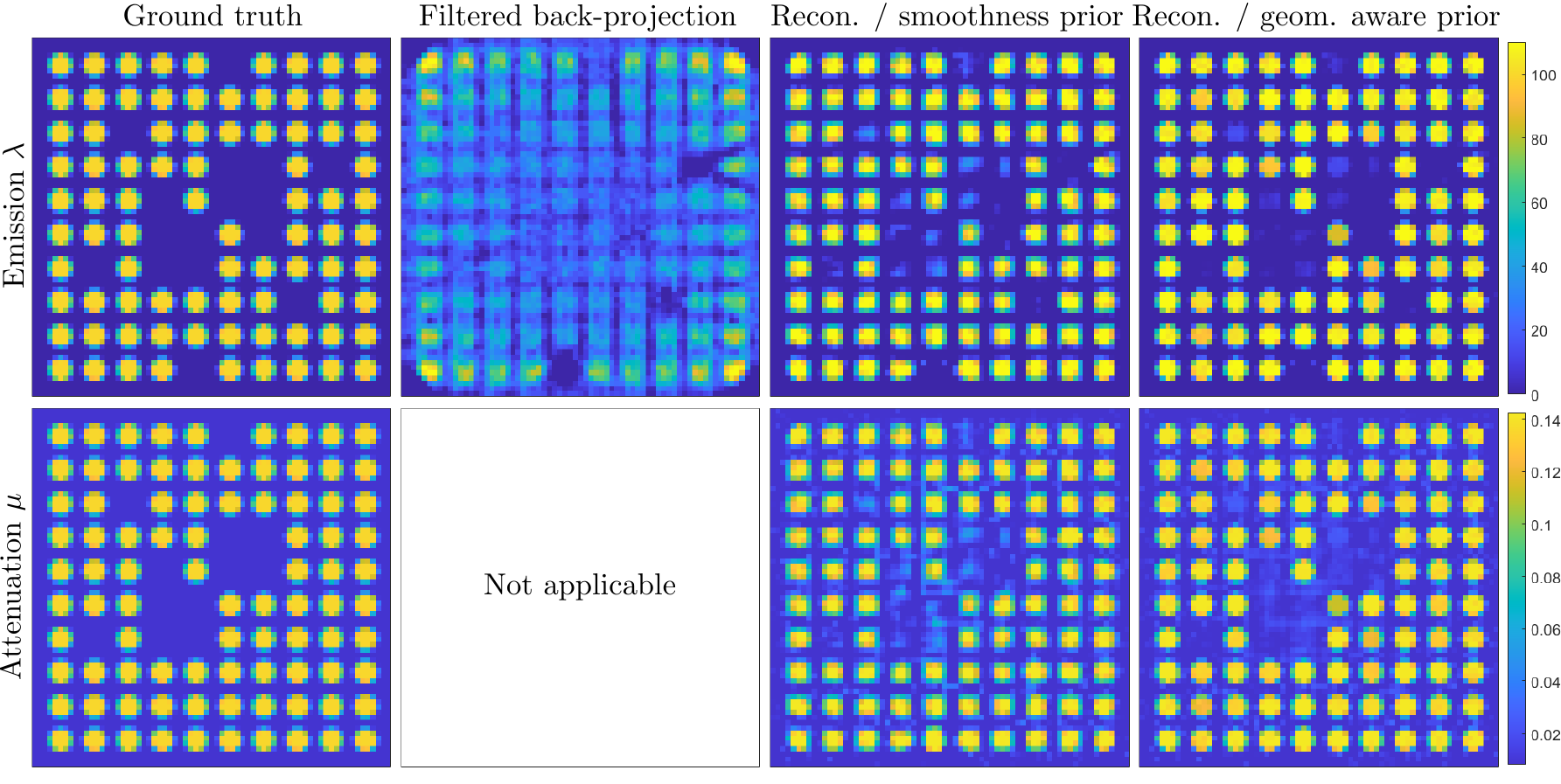}}

\captionsetup{width=\wideimwidth}
\caption{The ground truth and the reconstruction images cropped to include only the $69\times 69$ pixel area that includes the fuel assembly. In the top row there are the emission images and in the bottom row the attenuation images. In columns from left to right: ground truth, the FBP reconstruction, the iterative reconstruction using the smoothness prior, and the same using the geometry aware prior.}
\label{fig:recons}

\end{figure}

Cross-sections of the phantoms used in simulating the data, downsampled by a factor of 4 to the PGET intrument's reconstruction resolution of $165\times165$ and cropped to include only the $69\times 69$ pixel area of the fuel assembly, can be seen in Figure~\ref{fig:recons}. At this resolution, one pixel represent an area of $2\text{\ mm}\times 2\text{\ mm}$ when compared with the PGET instrument size.

The phantoms depict a GE12 assembly in water very near the center of the tomograph. An unmodified GE12 assembly consists of 92 UO\textsubscript{2} rods with a diameter of 8.8~mm on a $10\times10$ lattice with two $2\times2$ regions without fuel. We modified this nominal GE12 assembly to include both missing rods and rods replaced by fresh UO\textsubscript{2} rods at varying distances from the assembly center. The missing rods are in the top left half of the assembly and the replaced rods are in the lower right half. The attenuation phantom does not include the steel support structures of the assembly nor the steel interior wall of the tomograph. The attenuation values used are those corresponding to the 662~keV gamma-rays emitted by \textsuperscript{137}Cs, namely, 0.0085~mm\textsuperscript{-1} for water and 0.1356~mm\textsuperscript{-1} for UO\textsubscript{2}. 

We compare reconstructions done using the two different regularization terms, the smoothness prior and the geometry aware prior, both using the same bounds, and also a reconstruction done using FBP. All the reconstructions used 90 measurement angles spaced evenly over the full 360-degree rotation.

The upper and lower bounds for emission and attenuation values that determine the triangle of the linear bounds used in the iterative reconstructions were slightly extended from the minimum and maximum values used in simulating the data. Namely, the emission upper bound was 10\% larger than the value used for UO\textsubscript{2} in simulating the data and the lower bound was 0; the attenuation upper bound was 5\% larger than the value used for the UO\textsubscript{2}; and the attenuation lower bound was 5\% smaller that the value used for water.

The initial guess for the iterative reconstructions consisted of only water everywhere. The regularization parameters $\alpha_{\lambda}$ and $\alpha_{\mu}$ were chosen heuristically by sampling several values and choosing the ones yielding the best reconstructions according to the quality metrics in Table~\ref{table:metrics}.

\begin{table}

\centering
\includegraphics[width=\midimwidth]{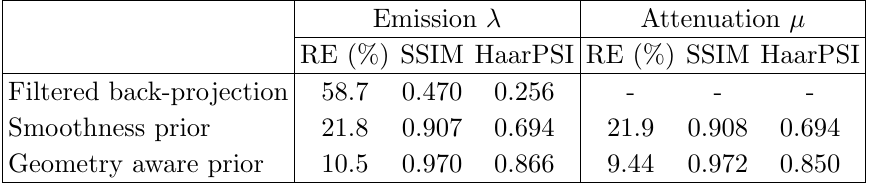}

\captionsetup{width=\midimwidth}
\caption{Metrics comparing the reconstruction to the ground truth: relative error (RE), structural similarity index (SSIM) and Haar wavelet-based perceptual similarity index (HaarPSI)}
\label{table:metrics}
\end{table}

The iterative reconstruction algorithm was stopped when the decrease in the objective function being minimized (the function in (\ref{eq:minprob})) dropped below 0.1\% between iterations. This resulted in 9 and 10 iterations for the smoothness and the geometry aware priors, respectively. The reconstruction process took 6 and 7 minutes, respectively, using a Matlab R2018a implementation of the algorithm on a laptop with i5-5300U CPU at 2.3 GHz and 16 GB of RAM.

Reconstruction images, cropped to include only the fuel assembly, can be seen in Figure~\ref{fig:recons}, and in Table~\ref{table:metrics} are values of different metrics comparing these cropped reconstructions to the cropped ground truth. The metrics used are the relative error (RE), computed as
\begin{align}
\frac{\left\Vert x_{\text{truth}} - x_{\text{recon}}\right\Vert_2}{\left\Vert x_{\text{truth}} \right\Vert_2}\cdot 100\%,
\end{align}
the structural similarity index (SSIM) \cite{Wang2004} and the Haar wavelet-based perceptual similarity index (HaarPSI) \cite{Reisenhofer2018}. For the first metric, smaller is better. The values of the SSIM and HaarPSI metrics range from 0 to 1 and larger is better.

The \textit{tout court} FBP reconstructed image contains pixels with negative values and its scale is entirely different from the ground truth. To better compare the methods, the FBP image is modified before applying the metrics and Figure~\ref{fig:recons} shows the modified version of the image. First, the negative values in the FBP image are set to zero, then the image is scaled so that the average of the pixel values in the cropped section matches the average of the cropped ground truth emission image.  Finally, the pixel values that are larger than the emission upper bound that was used in the iterative reconstruction are set to the value of the upper bound.

\begin{figure}[!t]

\centerline{\includegraphics[width=\wideimwidth]{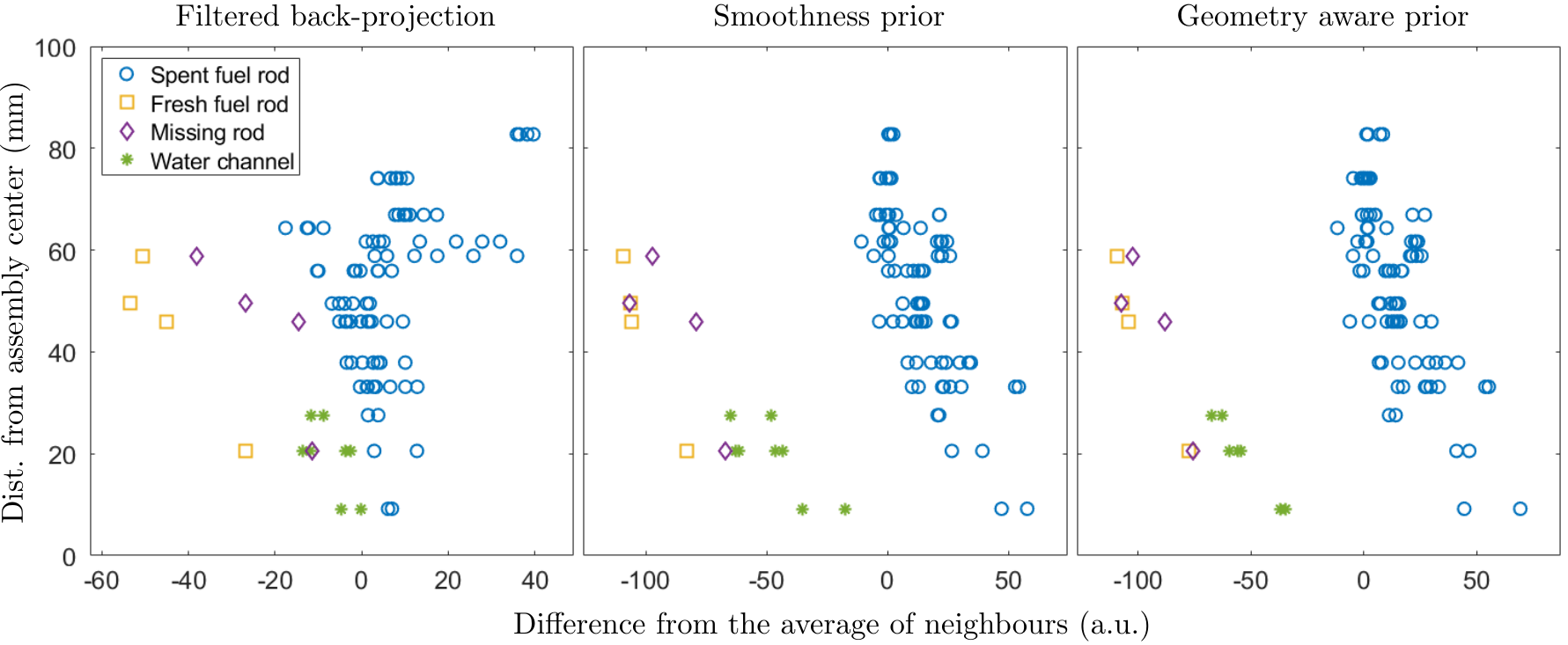}}

\captionsetup{width=\wideimwidth}
\caption{The difference of the emission value of a rod position from the average value of its neighboring positions plotted against the distance of the position from the assembly center. From left to right: FBP reconstruction, iterative reconstruction using the smoothness prior, and iterative reconstruction using the geometry aware prior.}

\label{fig:iaea}

\end{figure}

Figure~\ref{fig:iaea} shows plots constructed like those the IAEA uses for the classification of spent fuel rods and missing rods \cite{White2018}. For every fuel array position in the reconstructed emission image, the average value over the central $2\times 2$ pixels is computed to represent the emission value of that position. In the plots, the difference between a position's emission value and the average value of its neighbors is plotted against the distance of the position from the assembly center. The location of the water channels is not assumed to be known here, that is, they are not excluded when computing the average of the neighboring positions.

\begin{figure}[!t]

\def\imwidth{12cm}

\centerline{\includegraphics[width=\imwidth]{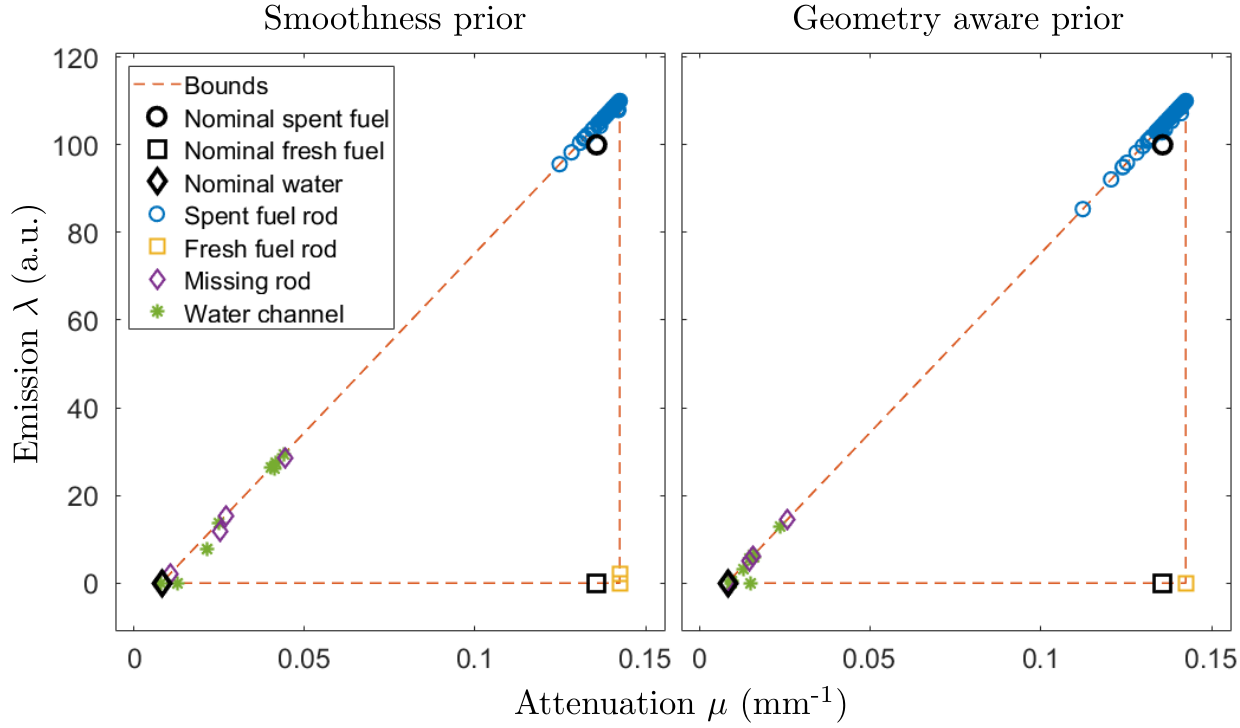}}

\captionsetup{width=\imwidth}
\caption{The emission and attenuation values of each rod position plotted in the emission-attenuation-plane for the iterative reconstructions using the smoothness prior (left) and the geometry aware prior (right).}
\label{fig:att-em}

\end{figure}

In Figure~\ref{fig:att-em}, emission and attenuation values of rod positions, computed again as averages of the $2\times 2$ pixel centers, are plotted in the emission-attenuation-plane. These plots present an alternate classification tool for fuel array positions in an SFA. However, they require that both emission and attenuation are reconstructed and hence are not applicable to the FBP reconstruction.

\section{Discussion}
\label{sec:discussion}

By all three metrics in Table~\ref{table:metrics} and by visual comparison of the reconstruction images in Figure~\ref{fig:recons}, the iterative reconstruction methods proposed here produced a significant improvement over the FBP reconstruction. Again by all three metrics, of the two regularization choices for the iterative method, the geometry aware prior produced better results than the smoothness prior, although the difference is not as significant as between the FBP and the iterative methods.

From the plots in Figure~\ref{fig:iaea} it is clear that compared to FBP the iterative reconstruction methods produce a better separation between the rod positions with and without spent fuel. Therefore, the proposed methods should allow for easier classification of the rods, if the classification is based on similar images. In this comparison, both choices for the prior term produce equally good results.

Reconstructing the attenuation and emission simultaneously with an iterative method offers two advantages. First, the attenuation correction makes the relative emission values more accurate. Second, the rod positions can be classified using the \textit{combination} of their emission and attenuation values, that is, based on the plots in Figure~\ref{fig:att-em}. Here, both iterative methods produce a clear separation between positions with spent fuel rods, fresh fuel rods and water. The geometry aware prior produces a somewhat tighter grouping of the positions with water in them than the smoothness prior.

The smoothness prior is a trade-off between model accuracy, computational performance and the amount of prior information used. Both the emission and attenuation values make a jump at the boundaries of rod and water, especially for higher reconstruction resolution. Therefore, assuming smoothness of those coefficients is not entirely accurate. However, the smoothness prior is well-understood, computationally efficient, and simple to implement and explain. Moreover, it does not assume any information about the geometry of the assembly. These are strong benefits in a real-world safeguards imaging task.

The geometry aware prior maintains the computational performance of the smoothness prior and improves on the model accuracy at the cost of making assumptions about the fuel assembly geometry. However, these assumptions are not unreasonable as they amount to identifying the fuel assembly type and its position, which is something that can be done from an FBP reconstruction. Such an identification is part of the current method used by IAEA \cite{White2017}. This identification is also very relevant for the second objective of GET, which is the quantitative assessment of individual fuel rod properties (e.g. the activity of key isotopes, cooling time, relative burnup), as knowledge of basic fuel parameters (e.g., assembly type and nominal fuel composition) is considered necessary for it \cite{Smith2016, JacobssonSvard2017}.

The bounds deliver a significant part of the reconstruction quality of the proposed method. They require some way of estimating the upper and lower bounds for the attenuation values and the upper bound for the emission values. The reconstruction quality is quite sensitive to getting these estimates somewhat correct and this is likely to be a challenge when moving to real data.

The attenuation bounds could be estimated based on the knowledge of the measurement energy window and assumptions about the materials being imaged, i.e., that water is the least and UO\textsubscript{2} the most attenuating material present. One way of estimating the emission upper bound could be to again identify the fuel assembly type and its location from an FBP reconstruction, and then quickly simulate a sinogram using the assumed attenuation values for water and UO\textsubscript{2} and some constant emission value for all the rods. The ratio between, e.g., the average value of the simulated sinogram and the constant emission value used in the simulation should be somewhat close to the ratio of the average value of the real sinogram and the upper bound for emission values, even if the real data comes from an assembly that is missing a few rods.

The way the regularization parameters $\alpha_{\lambda}$ and $\alpha_{\mu}$ were chosen here, by simply sampling several values and picking the ones that produced the best reconstruction, is time consuming. Yet these choices, once found, should work relatively well at least for some other reconstruction tasks. Since the forward model is linear in the emission $\lambda$, the input sinogram can be normalized so that neither the measurement time nor the intensity of the radiation should affect these choices much. Change of the assembly type or the measurement energy window(s) might have a larger effect. One may consider calibrating the regularization parameters using an imaging setup in a controlled environment before deploying the system to operative use. Another option is to use an automatic parameter choice method such as cross validation or discrepancy principle. The problem of parameter choice is outside the scope of this initial feasibility study.

The details of the current IAEA method are not publicly available. Although FBP is not the state-of-the-art reconstruction method for this application, it provides a well-known method for comparison. One could likely find a different way to scale the FBP reconstruction that would somewhat improve its performance by the metrics in Table~\ref{table:metrics}, but the shape of the plot in Figure~\ref{fig:iaea} is independent of this scaling. Also, using more measurement angles would enhance the FBP reconstruction to some extent, but the fact that the iterative method does not need more angles for a good quality reconstruction is a benefit as this allows for shorter measurement times.

With more optimized software and hardware, the reconstruction times would likely drop to a level that is acceptable from an operational point of view, i.e., below a minute or two.

This study is based on simulated data only. While we put effort into modelling the geometry of the measurement setup and the radiation physics, some real world phenomena are still left out. These effects include, for example, the fact that the detected radiation is not actually monoenergetic and that some part of the detected photons are scattered either in the fuel, in the instrument or in the detectors themselves. This will affect the accuracy of the forward model, but also the method by which the bounds are estimated, and will likely lead to further challenges when taking the method into practice. However, the computational model can be extended to include or approximate at least some of these additional features, so we believe that our regularized reconstruction approach offers a flexible framework for developing a real-world method that considerably improves image reconstruction in GET of SFAs. In particular, the reconstruction of an attenuation image of the SFA opens new possibilities for classification and inspection criteria for fuel.

\section{Conclusion}
\label{sec:conclusions}

In this paper, we propose an iterative reconstruction method for simultaneous reconstruction of emission and attenuation maps of axial cross-sections of SFAs from PGET measurements. The performance of the method, with two different regularization choices, was compared to FBP using simulated data with 90 measurement angles.

The proposed method shows significant improvement over FBP across multiple  metrics that compared the reconstructions to the ground truth.  It produces a better separation than FBP between spent fuel rods and rods that are missing or replaced with fresh fuel rods when classifying the rod positions in the emission reconstructions. Furthermore, the proposed method allows for a different, enhanced, approach to classification by using also the reconstructed attenuation information.

Of the two regularization choices for the proposed method, the geometry aware prior performs somewhat better than the smoothness prior, but the difference is not large. The geometry aware prior assumes some information about the geometry of the SFA being imaged, but this information can be estimated from an initial FBP reconstruction.

We expect further challenges in taking the method into practice with real data, but believe that this framework of regularized iterative reconstruction is a good starting point for a real world method that improves image reconstruction in GET of SFAs.

\section{Acknowledgment}
RB, CBC and PD were supported by Business Finland under Grant 1845/31/2014. RB also acknowledges partial support by the Academy of Finland through the Finnish Centre of Excellence in Inverse Modelling and Imaging 2018-2025.
TAB and SS acknowledge support by the Academy of Finland through the Finnish Centre of Excellence in Inverse Modelling and Imaging 2018-2025, decision number 312339, and Academy Project 310822. 
TH was supported by the Academy of Finland via the decision number 314879.

\bibliographystyle{ieeetr}
\bibliography{ms}

\begin{thebibliography}{10}

\bibitem{UN1970}
``Treaty on the non-proliferation of nuclear weapons,'' {\em United Nations
  Treaty Series}, vol.~729, no.~10485, 1970.

\bibitem{Levai1993}
F.~L\'{e}vai, S.~D\'{e}si, M.~Tarvainen, and R.~Arlt, ``Use of high energy
  gamma emission tomography for partial defect verification of spent fuel
  assemblies,'' Tech. Rep. STUK-YTO-TR 56, STUK, Helsinki, Finland, 1993.

\bibitem{White2017}
T.~{White}, M.~{Mayorov}, N.~{Deshmukh}, E.~{Miller}, L.~E. {Smith},
  J.~{Dahlberg}, and T.~{Honkamaa}, ``{SPECT} reconstruction and analysis for
  the inspection of spent nuclear fuel,'' in {\em Proc. 2017 IEEE NSS MIC},
  (Atlanta, GA, USA), 2017.

\bibitem{White2018}
T.~White, M.~Mayorov, A.~Lebrun, P.~Peura, T.~Honkamaa, J.~Dahlberg,
  J.~Keubler, V.~Ivanov, and A.~Turunen, ``Application of passive gamma
  emission tomography ({PGET}) for the verification of spent nuclear fuel,'' in
  {\em Proc. INMM 59th Annu. Meeting}, (Baltimore, MD, USA), 2018.

\bibitem{Mayorov2017}
M.~{Mayorov}, T.~{White}, A.~{Lebrun}, J.~{Brutscher}, J.~{Keubler},
  A.~{Birnbaum}, V.~{Ivanov}, T.~{Honkamaa}, P.~{Peura}, and J.~{Dahlberg},
  ``Gamma emission tomography for the inspection of spent nuclear fuel,'' in
  {\em Proc. 2017 IEEE NSS MIC}, (Atlanta, GA, USA), 2017.

\bibitem{BelangerChampagne2019}
C.~{B\'{e}langer-Champagne}, P.~{Peura}, P.~{Eerola}, T.~{Honkamaa},
  T.~{White}, M.~{Mayorov}, and P.~{Dendooven}, ``Effect of gamma-ray energy on
  image quality in passive gamma emission tomography of spent nuclear fuel,''
  {\em IEEE Transactions on Nuclear Science}, vol.~66, no.~1, pp.~487--496,
  2019.

\bibitem{Smith2016}
L.~E. Smith, S.~{Jacobsson-Sv\"{a}rd}, V.~Mozin, P.~Jansson, E.~Miller,
  T.~White, N.~Deshmukh, H.~Trellue, R.~Wittman, A.~Davour, S.~Grape,
  P.~Andersson, S.~Vaccaro, S.~Holcombe, and T.~Honkamaa, ``A viability study
  of gamma emission tomography for spent fuel verification: {JNT} 1955 phase
  {I} technical report,'' Tech. Rep. PNNL-25995, Pacific Northwest National
  Laboratory, Richland, WA, USA, 2016.

\bibitem{JacobssonSvard2017}
S.~{Jacobsson Sv\"{a}rd}, L.~E. Smith, T.~A. White, V.~Mozin, P.~Jansson,
  P.~Andersson, A.~Davour, S.~Grape, H.~Trellue, N.~Deshmukh, E.~A. Miller,
  R.~S. Wittman, T.~Honkamaa, S.~Vaccaro, and J.~Ely, ``Outcomes of the {JNT}
  1955 phase {I} viability study of gamma emission tomography for spent fuel
  verification,'' {\em ESARDA Bulletin}, vol.~55, 2017.

\bibitem{JacobssonSvard2015}
S.~Jacobsson~Sv{\"a}rd, S.~Holcombe, and S.~Grape, ``Applicability of a set of
  tomographic reconstruction algorithms for quantitative {SPECT} on irradiated
  nuclear fuel assemblies,'' {\em Nuclear Instruments and Methods in Physics
  Research Section A: Accelerators, Spectrometers, Detectors and Associated
  Equipment}, vol.~783, pp.~128--141, 2015.

\bibitem{Honkamaa2014}
T.~Honkamaa, F.~Levai, R.~Berndt, P.~Schwalbach, S.~Vaccaro, and A.~Turunen,
  ``A prototype for passive gamma emission tomography,'' in {\em Proc.
  Symposium on International Safeguards}, (Vienna, Austria), 2014.

\bibitem{Kunyansky2001}
L.~A. Kunyansky, ``A new {SPECT} reconstruction algorithm based on the
  {Novikov} explicit inversion formula,'' {\em Inverse Problems}, vol.~17,
  no.~2, pp.~293--306, 2001.

\bibitem{Censor1979}
Y.~Censor, D.~E.~Gustafson, A.~Lent, and H.~Tuy, ``A new approach to the
  emission computerized tomography problem: Simultaneous calculation of
  attenuation and activity coefficients,'' {\em IEEE Transactions on Nuclear
  Science}, vol.~26, pp.~2775--2779, 1979.

\bibitem{Berker2016}
Y.~Berker and Y.~Li, ``Attenuation correction in emission tomography using the
  emission data - {A} review,'' {\em Medical Physics}, vol.~43, no.~2,
  pp.~807--832, 2016.

\bibitem{Manglos1993}
S.~H. {Manglos} and T.~M. {Young}, ``Constrained {IntraSPECT} reconstruction
  from {SPECT} projections,'' in {\em 1993 IEEE NSS MIC Conf. Rec.}, vol.~3,
  (San Francisco, CA, USA), pp.~1605--1609, 1993.

\bibitem{Dicken1999}
V.~Dicken, ``A new approach towards simultaneous activity and attenuation
  reconstruction in emission tomography,'' {\em Inverse Problems}, vol.~15,
  no.~4, pp.~931--960, 1999.

\bibitem{Nuyts1999}
J.~Nuyts, P.~Dupont, S.~Stroobants, R.~Bennink, L.~Mortelmans, and P.~Suetens,
  ``Simultaneous maximum a posteriori reconstruction of attenuation and
  activity distributions from emission sinograms,'' {\em IEEE Transactions on
  Medical Imaging}, vol.~18, pp.~393--403, 1999.

\bibitem{Hawkins1999}
W.~G. Hawkins, C.-H. Tung, D.~Gagnon, and F.~Valentino, ``Some new sourceless
  and source-assisted attenuation correction methods for {SPECT} and {PET},''
  in {\em Proc. 1999 Int. Meeting on Fully Three-Dimensional Image
  Reconstruction in Radiology and Nuclear Medicine}, (Egmond aan Zee, The
  Netherlands), pp.~84--87, 1999.

\bibitem{Krol2001}
A.~{Krol}, J.~E. {Bowsher}, S.~H. {Manglos}, D.~H. {Feiglin}, M.~P. {Tornai},
  and F.~D. {Thomas}, ``An {EM} algorithm for estimating {SPECT} emission and
  transmission parameters from emission data only,'' {\em IEEE Transactions on
  Medical Imaging}, vol.~20, no.~3, pp.~218--232, 2001.

\bibitem{Gourion2002}
D.~Gourion, D.~Noll, P.~Gantet, A.~Celler, and J.~Esquerre, ``Attenuation
  correction using {SPECT} emission data only,'' {\em IEEE Transactions on
  Nuclear Science}, vol.~49, pp.~2172--2179, 2002.

\bibitem{Krol2002}
A.~Krol, I.~Echeruo, R.~B. Solgado, A.~S. Hardikar, J.~E. Bowsher, D.~H.
  Feiglin, F.~D. Thomas, E.~Lipson, and I.~L. Coman, ``{EM-IntraSPECT}
  algorithm with ordered subsets {(OSEMIS)} for nonuniform attenuation
  correction in cardiac imaging,'' in {\em Proc. SPIE 4684}, 2002.

\bibitem{Seppanen200}
A.~O. Sepp\"{a}nen, ``Correction of collimator blurring and attenuation in
  single photon emission computed tomography,'' m. sc. thesis, University of
  Kuopio, Kuopio, Finland, March 2000.

\bibitem{Thierry1999}
R.~Thierry, J.-L. Pettier, and L.~Desbat, ``Simultaneous compensation for
  attenuation, scatter and detector response for {2D} emission tomography on
  nuclear waste with reduced data,'' in {\em 1st World Congress on Industrial
  Process Tomography}, (Buxton, United Kingdom), pp.~542--551, 1999.

\bibitem{Mueller2012}
J.~L. Mueller and S.~Siltanen, {\em Linear and Nonlinear Inverse Problems with
  Practical Applications}.
\newblock Philadelphia, PA, USA: SIAM, 2012.

\bibitem{Schuster2012}
T.~Schuster, B.~Kaltenbacher, B.~Hofmann, and K.~S. Kazimierski, {\em
  Regularization methods in Banach spaces}.
\newblock Berlin, Boston: De Gruyter, 2012.

\bibitem{Kelley1999}
C.~T. Kelley, {\em Iterative Methods for Optimization}, ch.~3.2.3 and 3.3.5.
\newblock Philadelphia, PA, USA: SIAM, 1999.

\bibitem{Bonettini2009}
S.~Bonettini, R.~Zanella, and L.~Zanni, ``A scaled gradient projection method
  for constrained image deblurring,'' {\em Inverse Problems}, vol.~25,
  no.~015002, 2009.

\bibitem{Wang2004}
Z.~Wang, A.~C. {Bovik}, H.~R. {Sheikh}, and E.~P. {Simoncelli}, ``Image quality
  assessment: from error visibility to structural similarity,'' {\em IEEE
  Transactions on Image Processing}, vol.~13, no.~4, pp.~600--612, 2004.

\bibitem{Reisenhofer2018}
R.~Reisenhofer, S.~Bosse, G.~Kutyniok, and T.~Wiegand, ``A {Haar} wavelet-based
  perceptual similarity index for image quality assessment,'' {\em Signal
  Processing: Image Communication}, vol.~61, pp.~33--43, 2018.

\end{thebibliography}

\end{document}